# Development of a modular CdTe detector plane for gamma-ray burst detection below 100 keV

M. Ehanno[a], C. Amoros[a], D. Barret[a], K. Lacombe[a], R. Pons[a], G. Rouaix[a], O. Gevin[b], O. Limousin[b], F. Lugiez[b], A. Bardoux[c], A. Penquer[c]
(1) CESR, (2) CEA/DSM/DAPNIA, (3) CNES

[a]*Centre d'Etude Spatiale des Rayonnements, 9 avenue du Colonel Roche, 31028 Toulouse Cedex 04, France*
[b]*Commissariat à l'énergie Atomique, CEA Saclay DSM/DAPNIA/Service d'Astrophysique, bât. 709 L'Orme des Merisiers, 91191 Gif-sur-Yvette, France*
[c]*Centre National d'Etudes Spatiales, 18 Av. Edouard Belin, 31401 Toulouse Cedex 9, FRANCE*

**Abstract**

We report on the development of an innovative CdTe detector plane (DPIX) optimized for the detection and localization of gamma-ray bursts in the X-ray band (below 100 keV). DPIX is part of an R&D program funded by the French Space Agency (CNES). DPIX builds upon the heritage of the ISGRI instrument, currently operating with great success on the ESA INTEGRAL mission. DPIX is an assembly of 200 elementary modules (XRDPIX) equipped with 32 CdTe Schottky detectors (4x4 mm2, 1 mm thickness) produced by ACRORAD Co. LTD. in Japan. These detectors offer good energy response up to 100 keV. Each XRDPIX is readout by the very low noise front-end electronics chip IDeF-X, currently under development at CEA/DSM/DAPNIA. In this paper, we describe the design of XRDPIX, the main features of the IDeF-X chip, and will present preliminary results of the reading out of one CdTe Schottky detector by the IDeF-X V1.0 chip. A low-energy threshold around 2.7 keV has been measured. This is to be compared with the 12-15 keV threshold of the ISGRI-INTEGRAL and BAT-SWIFT instruments, which both use similar detector material.

## 1. Gamma-ray bursts

Gamma-ray bursts (GRBs) are the most violent phenomena observed in the Universe and occur at cosmological distances. They are well suited for probing the Universe to very high redshifts (possibly up to 10-15). The interstellar medium of GRB host-galaxies and the intergalactic medium can be probed by the GRB. GRB are likely to result from the gravitational collapse of a very massive star as suggested by the association of some GRBs with a particular class of supernovae of type Ic (SN Ic), sometimes called hypernovae. The GRB rate as a function of z becomes therefore a direct tracer of the star formation rate, in particular at high z. It becomes even possible, using GRB, to detect the first generation of stars (population III), which could have been particularly massive, and responsible for the re-ionization of the Universe and an early synthesis of metal elements. GRB are not only extreme events by the amount of released energy, but also by the physical processes at work. GRB, resulting from the faith of massive stars lead to the formation of a black hole in its center followed by relativistic ejection of matter. Particles accelerated in shocks taking place in the relativistic wind radiate in the gamma-rays. The afterglow emission, which manifests in a wider range of the electromagnetic spectrum results from the interactions of both jets with circumstellar and/or interstellar materials.

## 2. Requirements for a new generation GRB mission

Building upon the exciting SWIFT, INTEGRAL and HETE-2 results, a next generation GRB mission should in particular be able 1) to detect a large sample of GRBs independently of their durations and spectra, in particular X-ray rich GRBs expected to occur at large cosmological distances, 2) to observe GRBs simultaneously over the widest possible range of the electromagnetic spectrum, from optical, infra-red to X-rays and gamma-rays, 3) to provide GRBs positions with arcminute accuracy for follow-up observations with ground based telescopes.

The essential ingredient of such mission is therefore a wide field of view coded-mask imaging telescope with a good detection sensitivity from a few keV (3-4 keV) up to about 100 keV (Schanne et al. 2006)

## 3. On-going developments

To fulfill the requirements for a next generation GRB mission, as part of an R&D program funded by the French Space Agency (CNES), we have undertaken the development of an elementary detection module (XRDPIX), based on 32 segmented Cadmium Telluride (CdTe) semiconductor detectors provided by ACRORAD (Japan). Each CdTe cell is 4 mm × 4 mm × 1mm CdTe equipped with a Schottky contact at the anode and a guard ring at the cathode (e.g. Takahashi et al. 2002).

The 32 detectors are read out by the closely integrated IDEF X ASIC. The main challenge come from the fact that the low-energy threshold across the 32 detectors must remain as low as possible, with a goal of 4 keV on an XRDPIX. This implies on one hand that the ASIC must have an extremely low read out noise and on the other hand that the detector and ASIC hybridization is engineered to minimize the parasitic capacitance noise contribution (specified to be less than 2 pF).

The electrical and mechanical design of the XRDPIX therefore relies on an innovative approach. The high voltage (600 V) is supplied to the 32 detectors with a grid (in red on Figure 1) fixed at 9 positions, the 32 detectors are properly aligned with a tolerance of 100 microns and glued onto a high purity Al2O3 ceramic plate using a process that has been specifically elaborated for this task. The detector ceramic is to be stacked onto a second high purity ceramic plate (in yellow) supporting the ASIC. The proximity of the two ceramics minimizes the noise. This compact design is very robust to mechanical vibrations. It further provides the ability to operate on and test the two separate units. The hybridization of XRDPIX is on going at the moment (the first prototype are expected in the summer 2007). The choice of the CERAMTEC ceramic was motivated after intensive testing of other possible options.

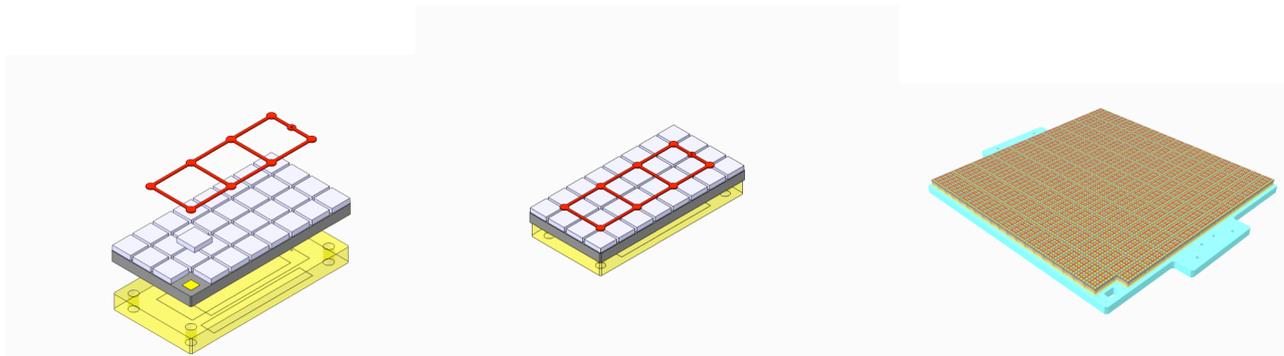

*Figure 1. Left) Schematic of the XRDPIX with the four ingredients shown separately: in red the grid providing the high voltages to the 32 detectors, in grey the ceramic plate with the 32 detectors, in yellow the second ceramic plate on which the ASIC chip is encapsulated. Middle) An XRDPIX assembled. Right) The DPIX detector plane (200 XRDPIX) mounted on its cold plate (in light blue).*

Taking advantage of the modular design of XRDPIX, it is easy to assemble a large area detector by staking properly as many XRDPIX as needed. An example of design for such an array is shown in Figure 1 where the individual XRDPIX are mounted onto a cold plate, allowing the detector to be operated at the desired temperature. The detector plane (called DPIX including its electronics) is an assembly of 200 elementary XRDPIX, which corresponds to an effective area of 1024 $cm^2$. It is currently foreseen as the detector for the SVOM/ECLAIRs X-ray/gamma-ray 2D coded mask imaging telescope (Shanne et al. 2006). The packaging of DPIX with its associated electronics is shown in Fig. 2.

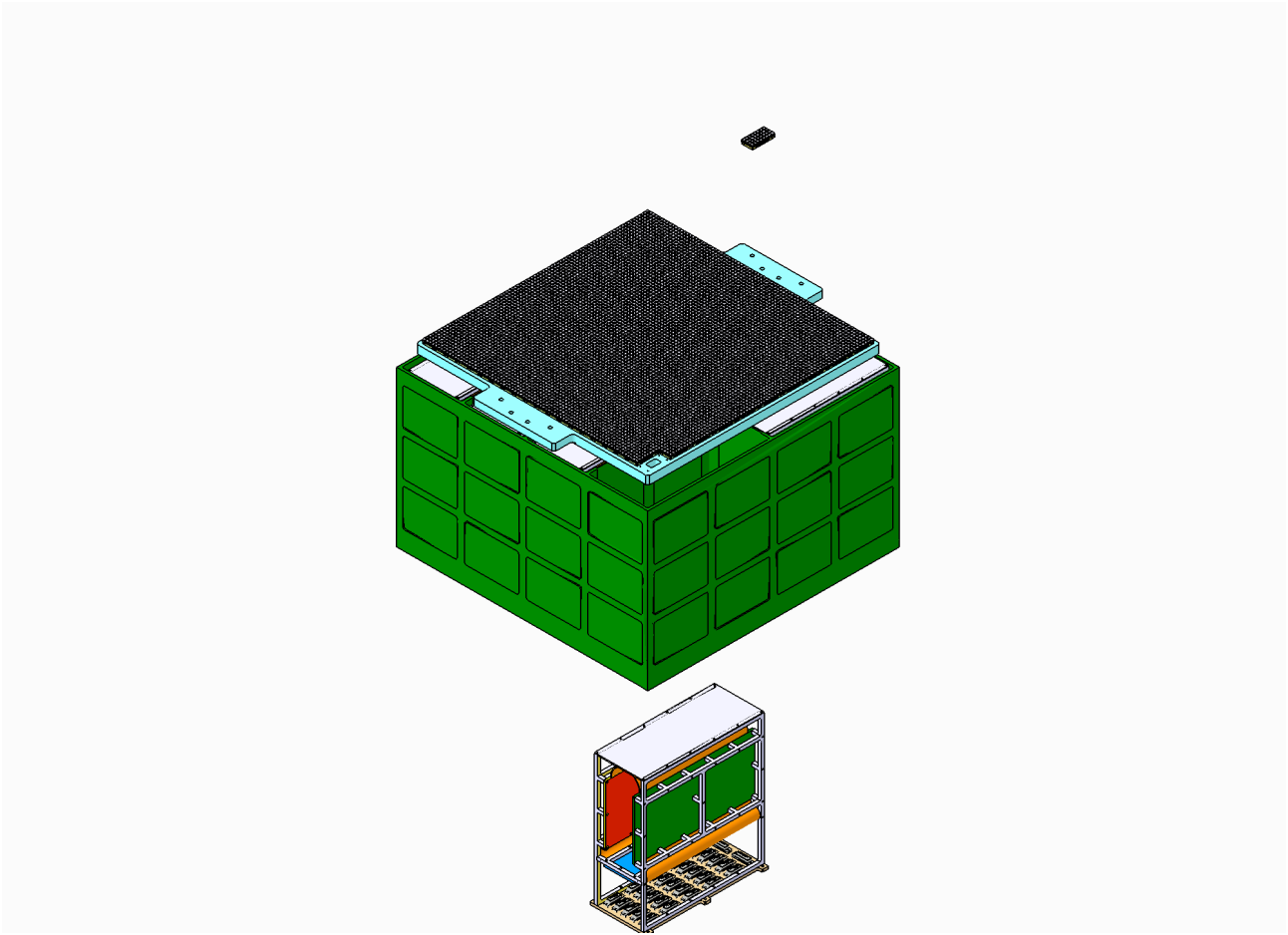

*Figure 2. Schematic of the DPIX detector plane made of 6400 CdTe detectors (200 XRDPIX, top), together with the mechanical structure holding DPIX and its associated readout electronic boxes (bottom)*

The central piece of XRDPIX is clearly the readout chip. IdeF-X is a very low noise multi channel integrated circuit. It is optimized for the readout of low capacitive (2-5pF) and low dark current (~ 1 pA to few nA) Cd(Zn)Te detectors or pixel arrays. The noise performance of this chip is very promising since the floor ENC was found to be 35 e- rms. It is processed with the standard AMS 0.35 μm CMOS technology. In its first version, each channel consists of a charge sensitive preamplifier, a pole zero cancellation stage, a variable peaking time filter and an output buffer (see schematic on Fig 3). The preliminary results indicate that a low-energy threshold of less than 4 keV is likely to be achieved with CdTe detectors (see below). The power dissipated is about 3 mW per channel (Limousin et al., 2005; Gevin et al., 2006). To fulfill the goal of a threshold of 4 keV implies that all the 32 detectors mounted on each XRDPIX are tested to ensure that the leakage current at operating temperature (-20 C as a baseline) remains within acceptable bounds.

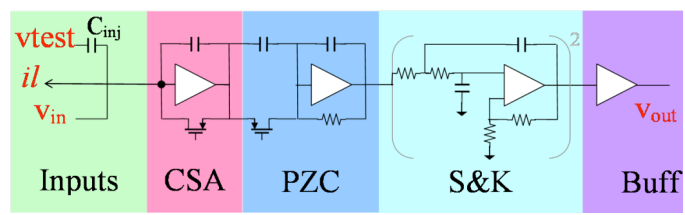

*Figure 3. Schematic of the IDEF-X ASIC (1 channel). A test input allows calibrated charge injection. A current input makes it possible to inject a leakage current in the channel. The injected charge is converted into voltage by the CSA. The output of the CSA is connected to the Sallen & Key type filters via a pole zero cancellation stage.*

## 4. Preliminary results

This Americium spectrum was recorded with a CdTe detector, cooled at -20C, polarized at 600 V, mounted on a ceramic plate and read out by the ASIC chip IDEFX1.0 (see Figure 4). **A low-energy threshold of 2.7 keV** and an energy resolution of 1.8 keV at 60 keV were measured, demonstrating that a threshold of 4 keV can be reached on 32 detectors properly tested and mounted on an XRDPIX.

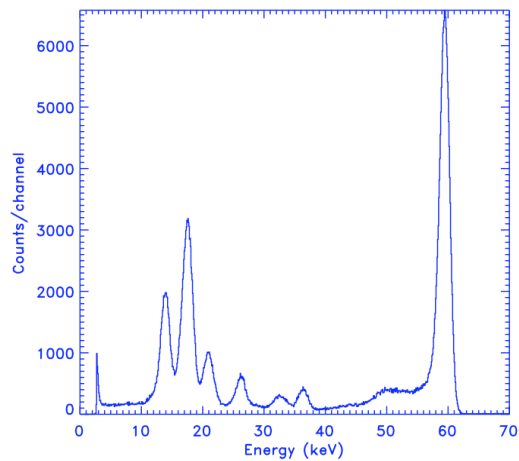

*Figure 4. The Americium spectrum obtained with IDEFX1.0 at -20C and polarized at 600 V. A low-energy threshold of 2.7 keV and an energy resolution of 1.8 keV at 60 keV were measured.*

## 5. Conclusions

Thanks to the support of the French Space Agency, we are finalizing the development of an array of 32 detectors of CdTe detectors at CESR, simultaneously with the conception of a very powerful low-noise low-power ASIC chip at CEA, in a collaborative approach with a common goal of building a large detector array with the lowest possible energy threshold. The goal of reaching a threshold as low as 4 keV seems within reach, matching the science requirements of a next generation gamma-ray burst detector, beyond those of currently operating missions, i.e. INTEGRAL (Lebrun et al. 2003) and SWIFT (Barthelmy et al. 2005) whose low-energy thresholds (around 15 keV) make difficult the detection of those very interesting GRBs with prominent emission in the X-ray band.